\documentclass{article}

\usepackage{preprint}
\usepackage{upgreek}				
\usepackage[utf8]{inputenc} 		
\usepackage[T1]{fontenc}    		
\usepackage[dvipsnames]{xcolor}	
\usepackage{hyperref}       		
\hypersetup{
    colorlinks = true,
    linkcolor = {NavyBlue},		
    anchorcolor = .,
    citecolor = {NavyBlue},		
    filecolor = .,				
    menucolor = .,
    runcolor = {cyan}, 
    urlcolor = {NavyBlue},		
    }
\usepackage{url}            
\usepackage{booktabs}       
\usepackage{amsfonts}       
\usepackage{nicefrac}       
\usepackage{microtype}      
\usepackage{fancyhdr}       
\usepackage{graphicx}       
\usepackage{setspace} 
\usepackage{soul} 
\usepackage{ragged2e} 
\usepackage[toc,page]{appendix}
\usepackage{longtable}
\usepackage{eso-pic}
\usepackage{tikz}
\usepackage[superscript]{cite}

\graphicspath{{./Figures/}}     

  \centering
\title{Correlation between Electrochemical Relaxations and Morphologies of Conducting Polymer Dendrites}

\author{
  Antoine Baron\textsuperscript{a}, Enrique H. Balaguera\textsuperscript{b}, S\'{e}bastien Pecqueur\textsuperscript{a}  \\
  \\
  a. IEMN, UMR 8520 \\
  Univ. Lille, CNRS, Univ. Polytechnique Hauts-de-France\\
  59000 Lille, France\\
  \\
  b. Escuela Superior de Ciencias Experimentales y Tecnolog\'{i}a, Universidad Rey Juan\\ Carlos, C/ Tulip\'{a}n, s/n, 28933 M\'{o}stoles, Madrid, Spain\\
  \\
  \texttt{sebastien.pecqueur@iemn.fr} \\
}

\begin{document}
\maketitle

\begin{abstract}
Conducting Polymer Dendrites (CPD) can engrave sophisticated patterns of electrical interconnects in their morphology with low-voltage spikes and few resources: they may unlock \textit{in operando} manufacturing functionalities for electronics using metamorphism conjointly with electron transport as part of the information processing. The relationship between structure and information transport remains unclear and hinders the exploitation of the versatility of their morphologies to store and process electrodynamic information. This study details the evolution of CPD's circuit parameters with their growth and shape. Through electrochemical impedance spectroscopy, multiple distributions of relaxation times are evidenced and evolve specifically upon growth. Correlations are established between dispersive capacitances of dendritic morphologies and growth duration, independently from exogenous physical variables: distance, evaporation or aging. Deviation of the anomalous capacitance from the conventional Debye dielectric relaxation can be programmed, as the growth controls the dispersion coefficient of the dendrite's constant-phase elements relaxation. These results suggest that the fading-memory time window of pseudo-capacitive interconnects can practically be conditioned using CPD morphogenesis as an \textit{in materio} learning mechanism. This study confirms the perspective of using electrochemistry for unconventional electronics, engraving information in the physics of conducting polymer objects, and storing information in their morphology, accessible by impedance spectral analysis.

\end{abstract}

\raggedright

\newpage 

\justifying

\section{Introduction}

Consumer-electronic technologies were so disruptive in the last decades that our lives have become greatly conditioned by our relationship with internet-of-things and artificial intelligence.\cite{Commission2023,Jaspers2022,Attie2022,Makarius2020} COVID’s pandemic has particularly demonstrated that organizational schemes heavily rely on electronics to promote remote activities, that demands will keep increasing, and that global economy strongly relies on electronic hardware manufacturing/retailing.\cite{Althaf2021,Belhadi2021,Xu2020,Armani2020} Its production is highly centralized and global, which has consequently generated lots of transportation to suppliers, with overcapacity preventing future shortages.\cite{Li2021,Silbermayr2014} Electronic hardware development is not exclusively driven by the cost of resources, but sizes its throughput on the global demand.\cite{IRDSexsum2022} Therefore, electronics' environmental footprint is not optimized at all: 
(1) Manufacturing is environmentally very expensive: circuits poorly value biomass and use precious or rare metals in their highest purity.\cite{Geng2023,Peng2023,Robinson2009} Pricings are not directly correlated with costs for manufacturing energy,\cite{Wang2023} nor materials to fabricate,\cite{Peng2023} nor water consumed by production facilities.\cite{Frost2019}
(2) Production is excessive and not personalized: downscaling electronic technologies secured overproducing electronic devices and their components over the years, highly standardized over batches with largely oversized capability features (memory, processors and sensors) in regards to a “standard user” demand.\cite{Triolo2024,Rapp2022,Waldrop2016,Rupp2011} 
(3) Items life-cycle is too low: they are frequently discarded when they are considered obsolete in regard to a newer version in the market, or they may be designed on purpose to diminish their lifespan and encourage consumption.\cite{MaitreEkern2016} Recycling is often more expensive than producing from raw materials.\cite{Balde2024,Chen2022,ShanthiBhavan2023}
Most of these issues related to resource consumption and pollution in electronics are to be correlated with the latency of the production/retailing circuit, the constant evolution of innovations and demands, and the inability of electronic hardware to physically adapt to personal needs of individual end-users during its lifespan. At the opposite of software applications which are always updated on hardware with minor impacts, conventional electronics is not generative, transformative nor transient.\cite{Jamshidi2022,Byun2020,Byun2019,Schmitt2018,Biswas2018,Biswas2016,Fu2016}\\
In nature, living organisms have this universal ability to adapt their structure to their environment over their life. They also multiply and evolve through many generations. 
Sessile organisms, having no ability to displace in space by themselves, physically extend and branch out, scavenging natural resources necessary for their development in their environment.\cite{Lee2021} Roots, mycellia, lichens, corals and slime molds grow on diverse substrates to adapt their topology and self-optimize to metamorphic structures, resilient to environmental changes. \cite{Tero2010,Simard1997}
Metamorphism is not necessarily intrinsic to growth (organisms may adapt by orientating, such as in the case of plants’ phototropism)\cite{Liscum2014}, but those having the ability to fabricate tissues and evolve in their environment are a true demonstration of the existence of physical mechanisms enabling \textit{in operando} hardware fabrication from environmental resources, which can be seen as an expression of their natural intelligence. 
The relationship between topological plasticity and computing functionality is also characteristic of our own brain, which activates the fabrication of new synapses and dendrites on specific areas depending on our activities and behaviors.\cite{Prigge2018,Kulkarni2012,Arikkath2012}\\
If the functioning of the brain has often drawn the inspiration to revisit electronic architectures for higher computing performances,\cite{Mehonic2022,Indiveri2011,Mead1990} topological materials compose only too few technologies for unconventional computing,\cite{Crepaldi2023,Marcucci2023,Adamatzky2010,Adamatzky2009,Cox1998} despite the fact that they may propose solutions other than increasing computing performances:\cite{Teo2023,Li2022,Jadoun2021,Tropp2021,Cramail2020,He2019,Kenry2018,Stavrinidou2017,Stavrinidou2015,Balint2014,Guo2013}
Indeed, conventional electronics does not self-repair, heal, multiply in ambient or compost. On the hypothesis that metamorphic electronics could be potentially diminishing the environmental footprint of electronic manufacturing by enabling it, one may question how concepts of computing with electrodes in or on a substrate material carrying both electricity and matter could condition the information transport. As physical mechanisms involving mass transport (such as convection and diffusion) are very distinct from electrostatic and electrodynamic principles governing holes and electrons in silicon, one may expect particularly different transport modes to be intrinsic to metamorphic junctions. Of prior importance, it is necessary to identify how mass-driven physical processes limit signal transport in a physically evolving junction.\\
In this study, we investigate the correlations between the physical impact of conducting polymer dendrites' (CPD) morphogenesis and the impedance between electrodes as a mechanism to program an electrical system. CPD are generated at low electric-field from the oxidation of monomers in electrolytes.\cite{Inagi2019,Koizumi2018,Watanabe2018,Ohira2017,Koizumi2016} They adopt various shapes according to the signal flowing between two electrodes.\cite{Eickenscheidt2019}. They can grow in water,\cite{Janzakova2021b,Janzakova2021a} and their pattern defines a specific conduction path for flowing electrons and charging ions.\cite{Janzakova2023,AkaiKasaya2020,Hagiwara2023} As organic mixed ionic-electronic conductors (OMIEC),\cite{Tropp2023,Paulsen2020} their behavior is more complex than metals:\cite{Janzakova2021,Cucchi2021,Petrauskas2021} CPDs carry both electrons and ions, so they behave as electrochemical transistors, with the latency of charges being ion-specific due to the material's varying affinity for different ions and molecules. In addition to the rich morphologies at different scales and the involvement of particles in their growth,\cite{Kumar2022,Janzakova2021a} a multi-physical origin of mass transport is expected: this study investigates on the link between transport and morphogenesis in the frequency domain by electrochemical impedance spectroscopy (EIS).\\

\newpage 

\section{Results and Discussion}

\subsection{Materials and methods}
Chemicals were purchased from Sigma Aldrich and have been used as such without further purification. Sodium polystyrene sulfonate (NaPSS) was used as an electrolyte at 1~mM concentration in deionized water. 3,4-ethylenedioxythiophene (EDOT) was used as a monomer undergoing oxidative electropolymerization: it was introduced at 10~mM prior solubilizing NaPSS as a surfactant. Parabenzoquinone was used equimolar as a proton scavenger and as an oxidizing agent to balance to electroneutrality upon electropolymerization (a mechanism is depicted in Fig.~\ref{fig:fig1}a,b). The growth was performed free-standing in the electrolyte drop on 25~$\upmu$m diameter gold wires cut from a spool purchased from Goodfellow. Every experiment uses freshly cut wires approximately 1~cm long. The plastic container is custom-made in order to prevent evaporation of volatile species in the electroactive solution, designed to hold 20~$\upmu$L of solution during up to one hour in ambient in a single experiment.\\

\subsection{Impedance characterization}
EIS was performed using a Solartron Analytical (Ametek) impedance analyzer in a 2-electrode setup, either intermittently or when both dendrites reach the state where the nearest distance between them is 40~$\upmu$m (a picture of the setup is shown in Fig.~\ref{fig:fig1}a,b).\\
Unless specified otherwise, the impedance spectra are obtained using ten samples per decade over the frequency range of 1~MHz to 100~mHz with an AC magnitude of 10~mV$_{rms}$. The DC bias is 0~V. The sampling rate and wave amplitude were set to minimize artefacts of voltage non-linearity and temporal drift. Except if mentioned otherwise, the time indicators on each label for impedance measurements define the total growth time of the CPD, rather than the practical duration of an experiment, which is characteristic of CPD aging.

\subsection{Circuit modeling}
Circuit modeling was achieved using ZView ver. 3.2b. The distribution of relaxation times (DRT) analysis was performed in Python using pyDRTTools\cite{Wan2015} and the following set of parameters:\\

\begin{table}[!h]
	\centering
	\begin{tabular}{c c c c}
	\textbf{Method of discretization}: & Gaussian & \textbf{Regularization derivative}: & 2\textsuperscript{nd} order \\ 
    \textbf{Parameter selection}: & mGCV & \textbf{Regularization parameter}: ($\lambda$) & 0.01\\ 	
    \textbf{RBF Shape Control}: & Shape Factor & \textbf{FWHM Control} & 1.0\\ 		
	\end{tabular}
\end{table}

\section{Results}

\subsection{Multiple Relaxations and Equivalent Circuit}

\begin{figure}[!h]
  \centering
  \includegraphics[width=1\columnwidth]{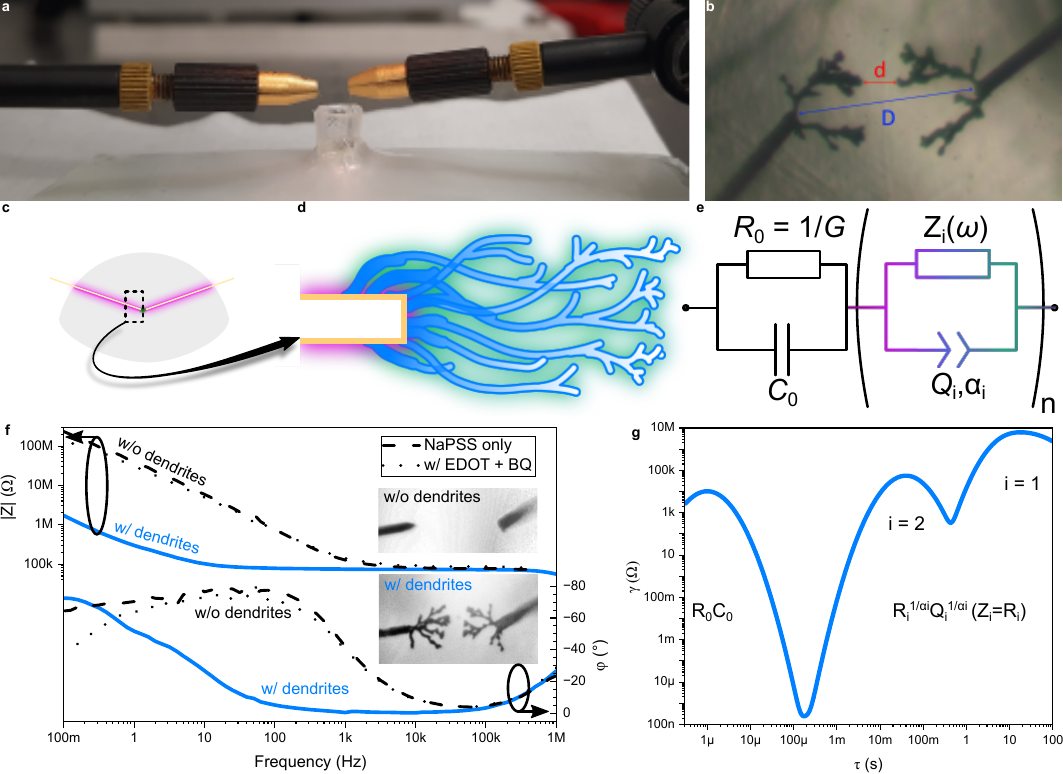}
  \caption{\textbf{Experimental setup and determination of an equivalent circuit model a,} Free-standing 2-electrodes setup to generate conducting polymer dendrites. Two micromanipulators are placed on the sides of the plastic container. Each supports one gold wire. The container is filled with 20~$\upmu$L of solution. A camera placed vertically above the container helps positioning the tips of the wires 240~$\upmu$m distant from each other. It also helps monitoring the reaction while it is taking place. \textbf{b,} Optical image of an incomplete dendritic connection. $d$ designates the nearest distance between the tips of the dendrites while $D$ is the distance between the tips of the gold wires. \textbf{c,} A schematic of the wires at the centimeter scale. The electrolyte is coloured in grey, while the surface of the wires in contact with the electrolyte is in purple. \textbf{d,} Zoomed-in representation of the tip of the wires, at the micrometer scale. The dendrites are drawn in blue, the surface corresponding to the interface between polymer and electrolyte is shown in green. \textbf{e,} The proposed equivalent circuit containing a RC element to modelize the electrolyte's contribution to the impedance and a variable number of ZARC elements to modelize the interfacial phenomena. \textit{Z}$_i(\omega)$ is intentionally left as generic as possible but $Z_i~=~R$ is considered for the fittings in this paper. \textbf{f,} Bode plots in magnitude and in phase of the uncoated wires in NaPSS only and in the full growth electrolyte (EDOT + NaPSS + BQ). In solid blue, Bode plot of the dendritic interconnection immersed in the growth electrolyte. The picture of the wires with and without electrolyte are also shown. \textbf{g,} Distribution of Relaxation Times of the dendritic interconnection shown in Fig.~\ref{fig:fig1}f.}
  \label{fig:fig1}
  \end{figure}

EIS was already performed on dendrites\cite{Scholaert2022,Janzakova2021,Janzakova2021a,Ghazal2021}. In the case of two gold wires supporting two distinct CPD, a clear non-Debye relaxation is observed.\cite{Janzakova2021,Ghazal2021}. Despite the fact that impedance is highly dependent on the CPD morphology grown at the apex of both gold electrodes, the influence of the wires on the signal propagation has not been established.\cite{Janzakova2021} Later, the impedance from CPD@Au to a naked silver wire showed a more ideal discharge with respect to Debye's relaxation, which highly depends on the voltage bias applied through the CPD during EIS:\cite{Scholaert2022} as PEDOT:PSS is a material which can be electrochemically doped or dedoped, the charge transport through the dendritic structure can affect the conductance of the polymer object, hence the charge distribution at the CPD/electrolyte interface. In our case, to analyze the properties of CPD with minimal dependence on the setup (such as voltage bias or the nature of the counter electrode), we chose to study the signal propagation from one dendrite to another without applying a voltage bias (V$_{DC}$ = 0). Fig.~\ref{fig:fig1}c,d show two wires used in the experiment with at their tip a dendrite. Both are immersed in a 20~$\mu$L volume of water contained in a plastic container. Different colors are used to distinguish the three existing interfaces between ions and electrons: a metal-electrolyte interface (purple), a polymer-electrolyte interface (green) and the OMIEC as heterojunction of ion-conducting semiconducting polymers (blue). 

Despite conducting polymers being low mobility semiconductors, the hole dynamics in PEDOT are faster than the ion drift mobility in water \cite{Stavrinidou2013}, so the slow dynamics of the ions dictate the overall dynamics of the system. As ions have different natures (size, charge, electro/nucleophilicity) and interfaces have different properties (roughness, surface, permittivity of the ion domain), it is expected that the junction limits signal propagation at different cut-off frequencies according to the different relaxations involved with the multiple ions and interfaces (Fig.~\ref{fig:fig1}e depicts a generic equivalent circuit assuming that all n+1 relaxations can be modeled in cascade form, one wire to the other). Fig.~\ref{fig:fig1}c,d illustrates the difference in scale of these interfaces and that dendrites only represent a small domain of the electrochemical system.
  In order to assess the independence of an impedance spectrum to the environment of the CPD (the wires that support them, and the redox species that surround them), three comparisons are presented in Fig.~\ref{fig:fig1}f on the same Bode plot. The impedance of a pair of gold wires immersed in an electrolyte containing only NaPSS\textsubscript{(aq)} is compared to its impedance after introducing EDOT\textsubscript{(aq)} and BQ\textsubscript{(aq)} (the total volume of electrolyte is reduced to the same volume prior the introduction of EDOT\textsubscript{(aq)} and BQ\textsubscript{(aq)} in NaPSS\textsubscript{(aq)}). A minor but noticeable contribution of EDOT and BQ can also be seen at low frequency in the spectrum. The magnitude remains too high for it to explain the increased conductivity after deposition. Thus, we can conclude that the interactions between EDOT, BQ and the gold wires do not limit the impedance of the system: nor by a faradaic process (electroplating \textit{in operando}) or by affecting the diffuse layer at electrode/electrolyte interfaces. The impedance of these same wires is compared to the one after electropolymerizing two CPD of comparable morphologies. The modification results in a drastic change of the impedance property of the system, by a decrease of at least two orders of magnitude of the impedance modulus (Fig.~\ref{fig:fig1}f). The resulting spectrum confirms that relaxation processes in polymer electrolytes can have characteristic time constants on the order of seconds.\cite{Kim2024,Li2023} Despite the fact that dendrites represent only a small fraction of the volume of the wires immersed in solution, CPD define most of the electrodes' impedance after growth, showing that the setup is suitable to study the impact of a CPD morphology on the impedance independently from the system. Another advantage of the free-standing wire setup is that no substrate is in contact with the electrodes: studies have shown that a CPD growth can be impacted by the affinity of a dendrite on a flat surface.\cite{Watanabe2018} As the phase of the signal appears to be rather complex, it is assumed that multiple relaxations rule the impedance modulus drop from 100~mHz to 1~kHz.\\
In order to assess it, the distribution of relaxation times (DRT) is performed assuming an elementary decomposition of the system in discrete Voigt elements in series (Fig.~\ref{fig:fig1}e,g). The range of the impedance characterization is enlarged from 1~MHz to 10~mHz to ensure that all relevant peaks can be fitted by the DRT. The distribution of relaxation times allows to visualize the number of processes taking place in the system without prior knowledge of its physics. Fig.~\ref{fig:fig1}g shows multiple relaxations. The highest frequency peak occurring at $\tau$(\textit{R}$_0$,\textit{C}$_0$) or $\tau_0$~	$\approx$~1~$\upmu$s could be attributed to the dipolar polarization of the solvent (we observe in general that this relaxation is invariant with the growth for any morphology, except when CPD merge and bridge one another). The other time constants are however very dependent on the growth conditions. They represent interfacial phenomena that are at this point difficult to interpret physically in the current stage, as the complex morphology and chemistry comes into play. It can be noted that the lowest frequencies processes are especially slow ($\tau_2$~$\approx$~30~s), which contrasts with the idea one might have for a RC system of such small dimensions (\textit{D}~=~240~$\upmu$m, \textit{d}~=~40~$\upmu$m). The peaks in the distribution are well spaced, distinguishable and symmetrical enough to follow a Cole-Cole distribution of two parameters $\alpha$ and $\tau$.\cite{Boukamp_2020} However, given the strong dependence of the result on the discretization parameters, fitting the peaks to a Cole-Cole distribution was not attempted. The result is shown in a log-log scale to visualize minor relaxations.  
  Based on the number of peaks displayed by the DRT, an equivalent circuit is defined in Fig.~\ref{fig:fig1}e depending on a number of n relaxations which compose the system: an ideal R|C relaxation at \textit{f}$_0$~=~1/2$\uppi\tau_0$ due to the geometric capacitance of the system (where the frequency is too high for ion motion and where water behaves as a dielectric) and numerous ZARC elements in a series, which in Fig.~\ref{fig:fig1}g n~=~2 modelize the longest relaxations. A preliminary fitting performed with this model (with \textit{Z}$_i$($\upomega$) as ohmic resistors) provides a satisfying match and suggests the following parameters:
  
\begin{table}[!h]
	\centering
	\begin{tabular}{|c|c|c|c|c|c|c|}
		\hline
		$1/G (k\Omega)$ & $R_1 (M\Omega)$ & $Q_1 (nF\cdot s^{\alpha_2-1})$ & $\alpha_1$ & $R_2 (k\Omega)$ & $Q_2 (nF\cdot s^{\alpha_1-1})$ & $\alpha_2$ \\
		\hline
		76.597 & 19.012 & 803.11 & 0.93028 & 121.520 & 829.83 & 0.84246\\
		\hline
	\end{tabular}
	\caption{Parameters for the dendrites grown in Fig. \ref{fig:fig1}}
\end{table}

The corresponding time constants are $\tau_1 = $ 18.73~s and $\tau_2 =$ 0.07~s, which are reasonably close to those obtained from the DRT. Considering the large difference in scale between both time constants, it is reasonable to assume that they are not to be attributed to one dendrite each, as their morphologies are comparable in size and features. The high values for \textit{R}$_1$ and $\tau_1$ indicate a long and highly resistive process, while the other ZARC element describes a relaxation several orders of magnitude faster, less resistive and also less ideal, suggesting that both relaxations originate from different physical processes, requiring deeper insight on varying the physical structure to understand the origin of the multiple processes.

\subsection{Elements Evolution upon Morphogenesis}

\begin{figure}[!h]
  \centering
  \includegraphics[width=1\columnwidth]{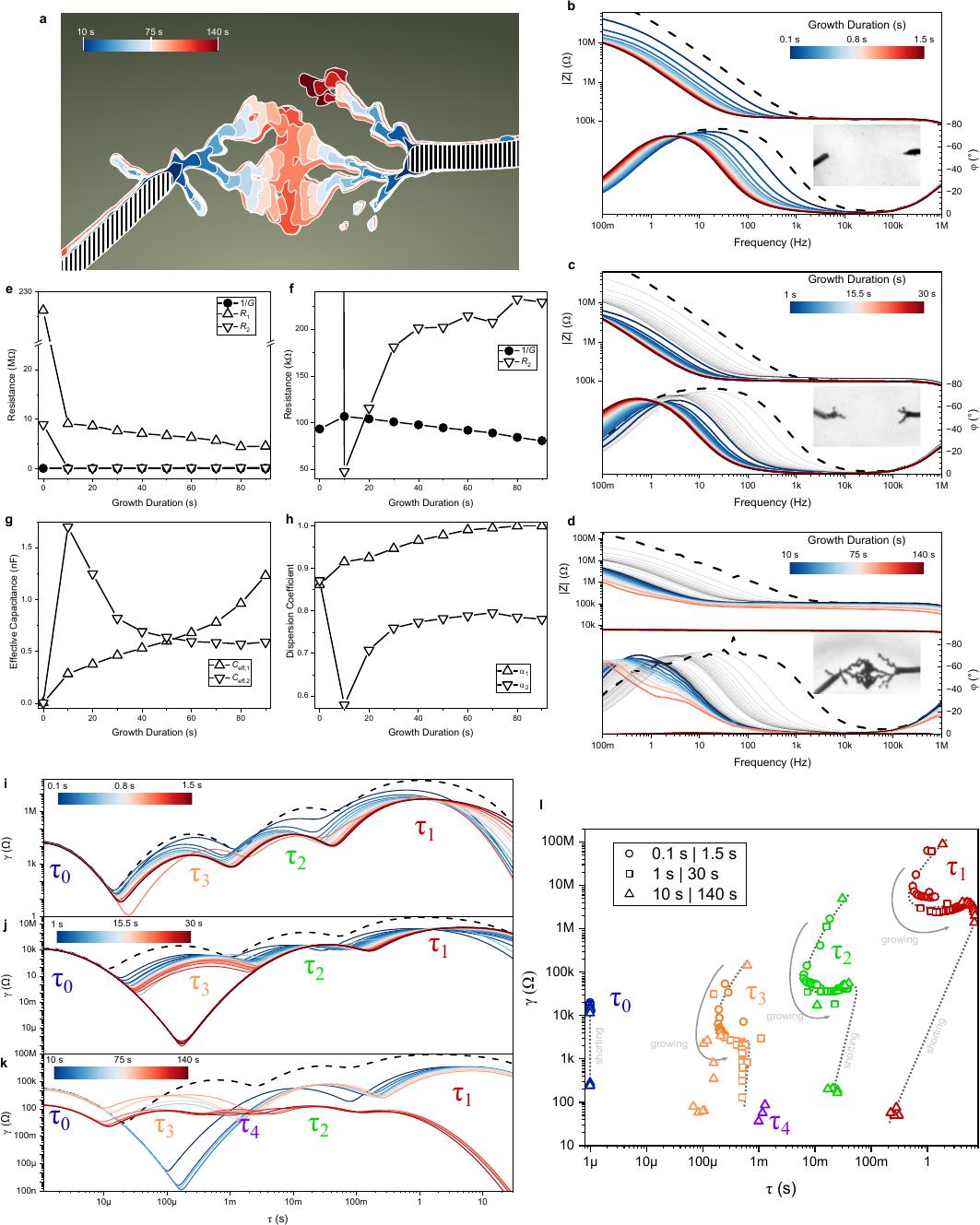}
  \caption{\textbf{Impedance measurements upon growths. a,} Vectorized-stacked images of time-lapsed pictures recorded during growth characterized in Fig.~\ref{fig:fig2}d. \textbf{b-d,} Bode plots in magnitude and in phase for dendrites grown from 0 to 1.5~s (b, \textit{T}$_{ep}$~=~0.1~s), from 0 to 30~s (c, \textit{T}$_{ep}$~=~1~s) and from 0 to 140~s (d, \textit{T}$_{ep}$~=~10~s). Insets: pictures of each deposition at the end of the experiment. In gray, combined impedance data of the earlier growths. \textbf{e-h,} Evolution of the fitted model parameters over growth, for the resistances (e,f), effective capacitances (g) and the dispersion coefficients (h). \textbf{i-k,} DRT of growths for \textit{T}$_{ep}$~=~0.1 (i), 1~s (j), and 10~s (k). \textbf{l,} Dependency of the DRT's log($\gamma$) local maxima with log($\tau$).} 
  \label{fig:fig2}
  \end{figure}

Fig.~\ref{fig:fig1}'s impedance is representative of disconnected symmetrical dendrites grown in standard conditions (100~Hz, 50\% duty-cycle). A number of time constants can be identified however, without information on the physical origin of the underlying processes. To understand this origin, the system is perturbed by several variables: the first one of interest is the stage completion under steady conditions of growth. To evaluate this effect, a set of EIS measurements is performed intermittently by interrupting a single growth at multiple stages, starting from uncoated Au wires until both dendrites interconnect. Previous tests in Fig.~\ref{fig:fig1} have confirmed that EIS is not invasive and does not modify the CPD when no bias is applied. Conventional EIS cannot be performed \textit{in operando} with the growth to characterize its impedance without faradaic processes, but intermittently interrupting a growth does not perturb the morphology by itself.\cite{Janzakova2021b} Such interruptions are depicted in Fig.~\ref{fig:fig2}a where each white edge separating different colors corresponds to a time where impedance was measured while the system remained at rest. The choice for the \textit{T}$_{ep}$ sampling is crucial, accounting for the time required to characterize the system at each step (approximately one minute per EIS) while CPD growth appears harder to trigger over time. A first growth at \textit{T}$_{ep}$~=~10~s is performed to observe the EIS evolution from nucleation to interconnection with 14 samples (Fig.~\ref{fig:fig2}a,d) and shorter \textit{T}$_{ep}$ at 1~s (Fig.~\ref{fig:fig2}c) and 100~ms (Fig.~\ref{fig:fig2}b) were performed to observe the drastic change of impedance at the very beginning of a growth. Despite the three growths occurring under the same experimental conditions but leading to three distinct morphologies, a continuum is observed in the impedance decrease (Fig.~\ref{fig:fig2}b-d). It should be mentionned that \textit{T}$_{ep}$~=~1~s and \textit{T}$_{ep}$~=~0.1~s series are generated at higher 50~mV$_{rms}$ as the impedance is particularly high and noisy at this stage of completion. The increase in V$_{rms}$ during the intermittent EIS does not appear to affect the growth. It appears that most of the impedance change occurs at the very beginning of the growth: the very first milliseconds are where the magnitude in the low frequencies falls the most. The top of the phase also slightly drops and shifts toward the low frequencies.
At high frequency, most of the change occurs when the distance between the wires is reduced as they get closer to each other, lowering the electrolyte's resistance 1/\textit{G}. A plateau in the 1-10~Hz range is also gradually appearing. 
The data of the \textit{T}$_{ep}$~=~10~s set, strictly reduced to the part of the experiment where the dendrites are growing but are not yet connected, are fitted with the equivalent circuit defined previously. The fitted parameters are presented Fig.~\ref{fig:fig2}e-h. Two R|CPE elements are considered (subcircuits of a resistor parallel to a constant phase element), modeling the slowest relaxations. CPE$_{i}$’s impedance is given by Z$_{CPE_{i}}$($\upomega$)~=~1/\textit{Q}$_{i}$(j$\upomega$)$^{\alpha_{i}}$, where \textit{Q}$_{i}$ is a pseudo-capacitance, j$^{2}$~=~-1, $\upomega$ the angular frequency related to the frequency \textit{f} by $\upomega$~=~2$\uppi$\textit{f}, and $\alpha_{i}$ (0~<~$\alpha_{i}$~<~1) the dispersion coefficient.\cite{Gateman2022} For each Voigt circuit, an effective capacitance \textit{C}$_{eff,i}$ can be defined as: \textit{C}$_{eff,i}$~=~\textit{Q}$_{i}^{1/\alpha_{i}}$\textit{R}$_{i}^{1/\alpha_{i}-1}$.\cite{Hirschorn2010} Interestingly, while most of the system's resistances are decreasing upon electropolymerization including the electrolyte resistance 1/\textit{G}, \textit{R}$_{2}$ increases slightly. The opposite trends of \textit{R}$_{1}$ and \textit{R}$_{2}$ with the growth validate that both relaxations originate from different processes and are not specifically to be attributed to each dendrite distinctive morphology.\cite{Baron2024}
And while the dispersion coefficient $\alpha_{1}$ converges toward 1.0, becoming equivalent to an ideal capacitor, $\alpha_2$ quickly stabilizes around 0.8, although this behaviour is not systematically observed. 
The three DRT in Fig.~\ref{fig:fig2}i-k displays actually more relaxations than previously observed: a total of four time constants despite the noise being removed by the 50~mV AC magnitude, cementing the idea that the system has four relaxations taking place with or without dendrites, and that dendrites do not create additional relaxations but rather affect the existing ones, defined as R|C$_0$($\tau_0~\approx$~1~$\upmu$s) and R|CPE$_i$($\tau_i~\approx$~1-10~s for i~=~1, 10-100~ms for i~=~2 and 0.1-1~ms for i~=~3 - pointed by the blue marker for R|C$_0$ and respectively red, green and orange for R|CPE$_i$ in Fig.~\ref{fig:fig2}i-l). 
The shift toward the low frequencies previously observed in the Bode plot translates in the time domain to a shift toward the slow relaxations, as $\tau_2$ and $\tau_3$ get longer over time. Meanwhile, the third R|CPE$_3$ relaxation, also subjected to this shift, tends to disappear during the growth.
Finally, the high frequency relaxation R|C$_0$ is stable over time until both dendrites are connected, where the electrolyte's contribution to the impedance becomes minimal.
The local maxima of each relaxation in the DRT are plotted in Fig.~\ref{fig:fig2}l for all EIS. Again, data are consistent between the sets. R|CPE$_1$, R|CPE$_2$ and R|CPE$_3$ show the same type of trajectory in the plane, rapidly stabilizing at a certain $\gamma$ before their time constant gradually become longer, and finally dropping upon dendrites' completion. Only R|C$_0$ follows a straight vertical trajectory, unaffected by the dendrites' growth or the decreasing distance between the electrodes. We note that the initial value of $\tau_1$ is approximately 1~s and corresponds to the configuration without polymer on the wires, meaning that, already when charging at a gold interface, ions still need seconds to stabilize over a 240~$\upmu$m gap in water. This stresses the fact that the dynamics of the CPD are not only conditioned by their morphology but also by the dynamics of a given electrolyte in a given environment and that a given morphology may result in very different dynamics if solvents and ions are changing (which has not been investigated at the moment).
The DRT profiles of the system when dendrites are fully connected are also displaying four relaxations in total, each being of similarly low peak values. This contrasts with the previous DRT profiles where the $\gamma$ would increase with the time constants. In this configuration, no relaxation is longer than one second. An ionic contribution to the current likely still exist but is minor when compared to the electronic conduction once CPD are in contact with one another. Also, the growth itself seems interrupted, as the DRT spectra do not change substantially. However, some growth can still happen to incomplete side branches (often considered 'dead' branches, although it is a fact that they keep promoting growth). Their contribution to the impedance is overshadowed by the primary branches being fully connected, although not completely made silent on the EIS.
  
\subsection{Decoupled Effects of Time and Distance}

\begin{figure}[!h]
	\centering
	\includegraphics[width=1\columnwidth]{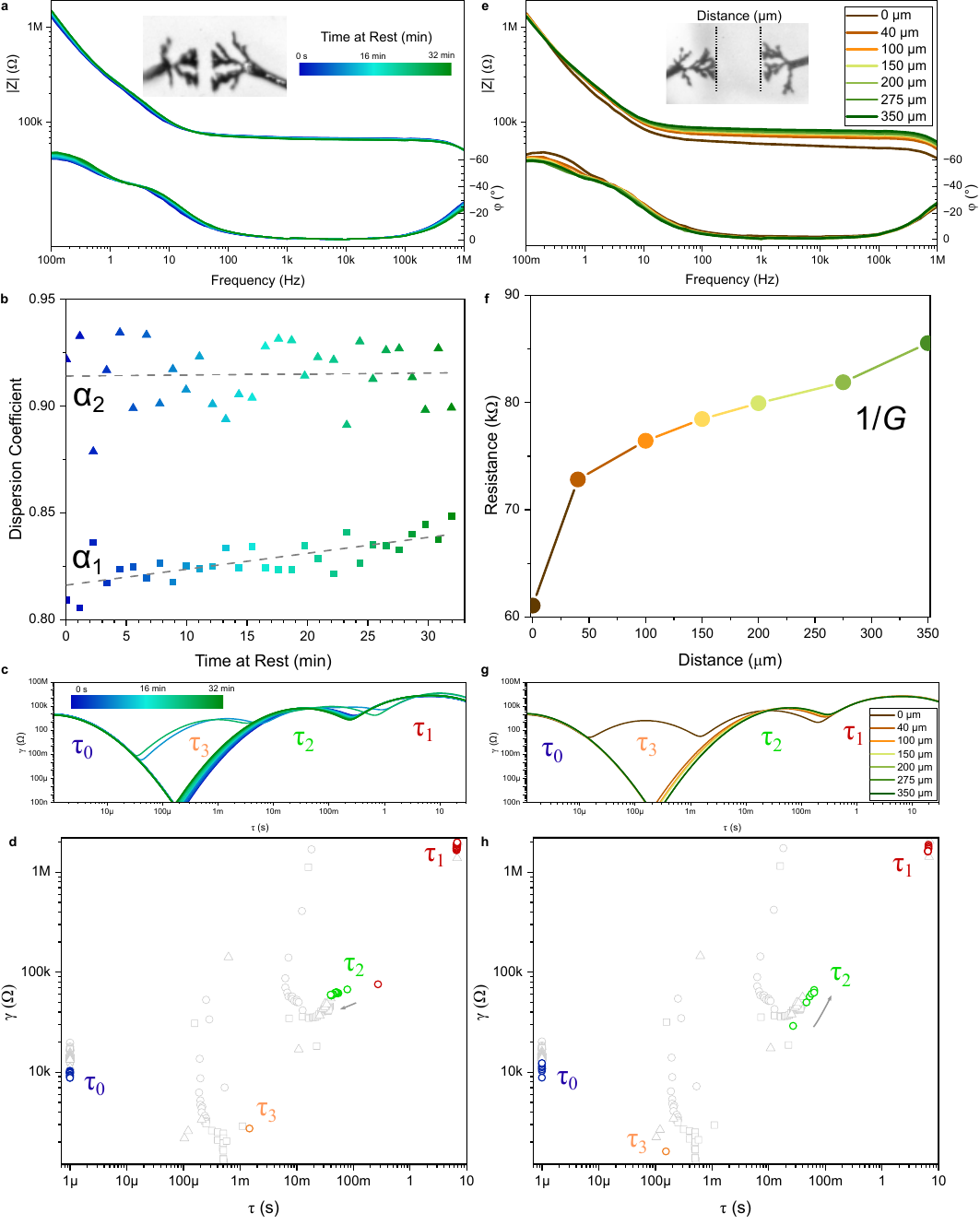}
	\caption{\textbf{Dendrite stability and impact of their distance on the impedance. a,} Bode plots in magnitude and in phase for 32~min (measured every minute). Inset: picture of the dendrites at the end of the experiment. \textbf{b,} Dispersion coefficients stability, from Fig.~\ref{fig:fig3}a's data fitting. \textbf{c,} DRT from Fig.~\ref{fig:fig3}a's data. \textbf{d,} Dependency of Fig.~\ref{fig:fig3}c's DRT local maxima of log($\gamma$) with log($\tau$) (compared to Fig.~\ref{fig:fig2}'s data in grey). \textbf{e,} Bode plots in magnitude and in phase for different distances. Inset: picture of the dendrites 100~$\upmu$m apart. \textbf{f,} Dependence of the fitted electrolyte resistance with the dendrite distance. \textbf{g,} DRT for the different gap widths. \textbf{h,} Dependence of Fig.~\ref{fig:fig3}g's DRT local maxima of log($\gamma$) with log($\tau$) (compared to Fig.~\ref{fig:fig2}'s data in grey).}
	\label{fig:fig3}
\end{figure}

Made of a conductive material, CPD decrease systematically the electrolyte gap as the material grows during a specific time. The effect that a given morphology has on the impedance shall generically depend on the morphology itself and not the distance between vicinal dendrites, nor the time dendrites age in an electrolyte. These two parameters have been studied separately.\\
To study the impact of aging, measurements have been performed on an incomplete symmetric morphology without modifying their structure nor their environment (Fig.~\ref{fig:fig3}a). EIS is performed approximately every minute over 32 minutes in solution, as the dendrites are left as is (minor manipulations of their position have been performed to compensate an apparent wire drift happening over time). At high frequencies, the impedance remains steady by the stability of its modulus and its phase. This indicates that water composing the electrolyte does not evaporate at a significant point and that the electrolyte concentration does not change noticeably as a result. An effect at low frequencies is however noticeable in the phase diagram, meaning that time has a minor impact on a dendrite impedance. Fig.~\ref{fig:fig3}b indicates that one of both relaxations becomes slightly more ideal overtime (however the average value of related dispersion coefficient $\alpha_2$ drifts by less than 0.01 over 13 minutes). On the DRT, R|CPE$_2$ relaxation trajectories in Fig.~\ref{fig:fig3}c,d is noticeably affected by aging, as $\tau_2$ drifts toward faster relaxation times. Possible reasons explaining the aging may be related to evaporation of EDOT and BQ in solution over time, affecting more rapidly their concentration than water's evaporation, and for which a minor effect of their presence on an EIS has been evidenced in Fig.~\ref{fig:fig1}f.\\
The impact of CPD distances have been studied independently from their morphology on the same growth of CPD pair by adjusting their distance via micro-manipulators supporting the gold wires (without pulling the CPD out of the growth electrolyte). The Bode plots in Fig.~\ref{fig:fig3}e show that a significant drop in magnitude exists in the highest frequency region (\textit{f}~>~10~Hz), likely caused by the decrease in electrolyte resistance since the distance that ions have to cross is significantly reduced (despite their trajectory from one morphology to another being complex). The fitted values in Fig.~\ref{fig:fig3}f confirms that the electrolyte resistance is the parameter responsible for the high frequency variation observed in the impedance spectra in Fig.~\ref{fig:fig3}e. At the particular case where \textit{d}~=0~$\upmu$m and the dendrites appear in contact with each other, the magnitude is slightly dropping while the frequency increases. This likely leads to the R|CPE$_3$ relaxation being visible in the DRT, unlike the other cases, as shown in Fig.~\ref{fig:fig3}g. As for the low frequencies, an effect on the phase is noticed. The trajectory of the R|CPE$_2$ relaxation, depicted in green in Fig.~\ref{fig:fig3}h, shows that R|CPE$_2$ is, again, the only relaxation affected by the varying gap for a given morphology. R|CPE$_1$ remains poorly affected by the CPD distance, which indicates that the underlying process is defined by the interface with the electrolyte on both electrodes rather than the volume of electrolyte between dendrites. It should be pointed also that because each measurement requires a specific duration for the acquisition, the set of measurements may be affected by the effect of the time has shown previously in Fig.~\ref{fig:fig3}a-d: it is, in fact, suspected that the slight drift at low frequency in Fig.~\ref{fig:fig3}e is rather due to aging rather than distance. \\
Among the multiple relaxations taking place in the system, only R|CPE$_2$ is yielded by a mechanism which is moderately affected by time, drifting the dispersion coefficient by less than 0.01 over 13~minutes. Distance has no effect on the impedance at the lower frequency part of the spectra where electrochemical relaxations dominate the dynamics.\\

\subsection{Impact of Dendrite's Morphology}

\begin{figure}[!h]
  \centering
  \includegraphics[width=0.96\columnwidth]{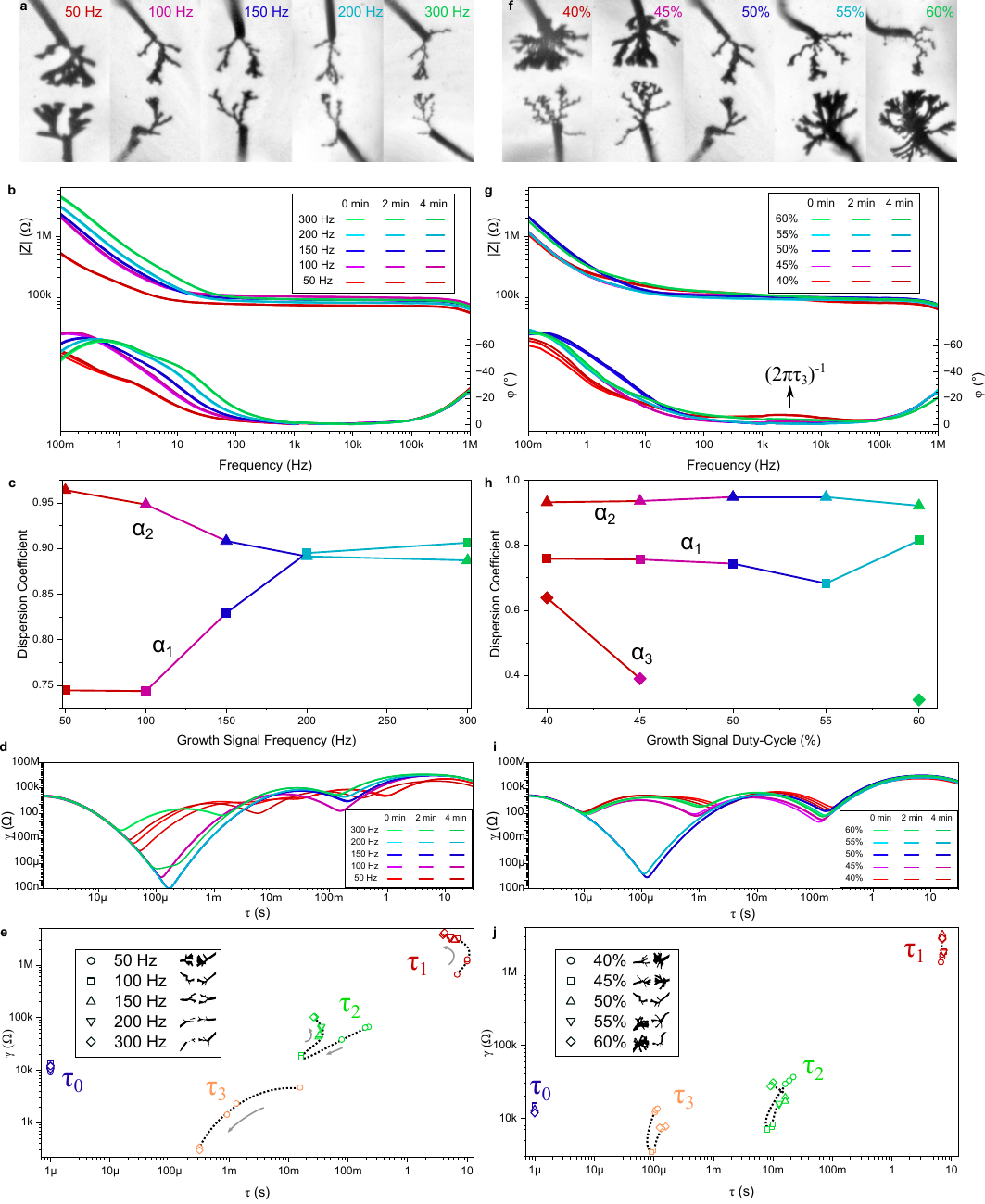}
  \caption{\textbf{Different dendrites' morphologies and their respective impedance a,} Pictures of the different morphologies obtained by varying the frequency of the square wave voltage. \textbf{b,} Bode plots of the corresponding dendrites, each impedance taken three times in a row to evaluate the stability over time. \textbf{c,} $\alpha_1$ and $\alpha_2$ extracted from the last impedance data of the series of three, for each dendrite. \textbf{d,} Corresponding DRT (including the series of three for each dendrite). \textbf{e,} Dependency of Fig.~\ref{fig:fig4}d's DRT local maxima of log($\gamma$) with log($\tau$). \textbf{f,} Pictures of the different morphologies obtained by varying the duty-cycle of the square wave voltage. \textbf{g,} Bode plots in magnitude and in phase for dendrites formed with different duty-cycles; again by sets of three measurements. \textbf{h,} $\alpha_1$, $\alpha_2$ and $\alpha_3$ extracted from the last impedance data of the series of three, for each dendrite. \textbf{i,} Corresponding DRT (including the series of three for each dendrite). \textbf{j,} Dependency of Fig.~\ref{fig:fig4}i's DRT local maxima of log($\gamma$) with log($\tau$).}
  \label{fig:fig4}
  \end{figure}

As the setup allows monitoring changes in a CPD impedance (Fig.~\ref{fig:fig2}) with a minor influence of the gold electrodes surface and electroactive electrolyte components (Fig.~\ref{fig:fig1}) or of the wire distance and solution aging (Fig.~\ref{fig:fig3}), different morphologies of CPD growth on pairs of different Au wires have been characterized to observe an impact of their fractality on the interconnection impedance (considering that different Au wires or CPD gaps would have no visible impact on the impedance).
Former studies showed that varying pulse signal frequency or duty-cycle during growth affects greatly the CPD morphology.\cite{Janzakova2021,Janzakova2021b}
Dendrites were formed with five different pulse frequencies (50, 100, 150, 200, and 300~Hz) and their impedance was measured at \textit{d}~=~40~$\upmu$m apart from each other. A constant delay of six minutes was introduced between adding the water in the container and taking the impedance measurement, so the drift effect is comparable from one morphology to another one. Additionally, each morphology was characterized three times at specific durations for fair comparison of their respective differences in regard to the temporal drifts (Fig.~\ref{fig:fig4}).
The resulting morphologies and Bode plots are presented respectively in Fig.~\ref{fig:fig4}a and Fig.~\ref{fig:fig4}b. The magnitude at high frequency does not exhibit a clear relationship with the frequency: it is assumed that for comparable distances \textit{d}, the electrolyte conductance is rather dominated by the number of branches on each CPD rather than their respective thickness (this feature is less controllable than the branch thickness, cheangeable by varying the growth frequency)\cite{Janzakova2021b}. The magnitude at low frequency follows a monotonous trend with the growth frequency: the higher the growth frequency, the lower the effective capacitance appears to be. This is to be attributed to the lower volume of OMIEC on thinner dendrites (grown at higher frequencies) than thicker ones (grown at lower frequencies). On the other hand, thicker dendrite exhibit lower and flatter impedance modulus and a trend in the phase which deviates even more than the ideal -90° Debye relaxation capacitor. From the impedance data fitting (Fig.~\ref{fig:fig4}c), the dispersion coefficients $\alpha_1$ and $\alpha_2$, initially distant from each other of about 0.2, not evolving much between 50 and 100~Hz, were shown to abrutptly converge between 100 and 200~Hz toward the same value of 0.896$\pm$0.001. Incidentally, most of the morphological changes also happen below 150~Hz. As the frequency increases, dendrites become less distinguishable from their morphology, same as their respective dispersion coefficient.
The DRT in Fig.~\ref{fig:fig4}d shows a monotonic evolution of two to three R|CPE$_i$ relaxations with the growth frequency. As dendrites get thinner, all relaxation time constants $\tau_i$ increase. This is in line with the idea that thicker dendrites may retain longer ions in their nano-structure as and OMIEC. The local maxima of the DRT in Fig.~\ref{fig:fig4}e show that thinner dendrites have higher values of $\gamma$ and $\tau_1$ overall, however the interpretation of the differences in spectra here is made difficult because of the relative existence of the third relaxation, which can cause a shift in the DRT and complicate the comparison.
Another effect that must be taken into account is that electropolymerization takes longer at high frequency for comparable voltage applied during the growth. A longer time for electropolymerization may lead to more accumulation of charged species within the bulk of the polymer, possibly leading to OMIEC with differenced of conductivity either for the ions and/or for the holes.
To verify that the relaxations are not specifically localized on one particular CPD compared to the other, dendrites are formed with different duty-cycle values on the pulse waveform used for their growth. Different polymerization times were required depending on the duty-cycle to reach comparable completion \textit{d}/\textit{D}, leading to a clear unbalance in their morphologies as shown in Fig.~\ref{fig:fig4}e. Branches that arise do not only grow on the tip of the gold wire, but also on its edge, as an unbalance in polymerization times gives more time to electroactive oligomer particles to reach longer distances.\cite{Kumar2022} On the other hand, the electrode with shorter electropolymerization times shows thinner branches, with regularly spaced sharp angles. Both electrodes cover a larger area with multiple growth directions, which seem to follow the electric field lines. Five different values of duty-cycle (40\%, 45\%, 50\%, 55\%, 60\%) were tested. Above 60\%, bubbles started appearing at the work frequency, disturbing the impedance measurement. Electrolyte bubbling occurs because the duty cycle affects the DC component of the periodic signal waveform. If the duty cycle deviates too much from 50\%, it can push the DC component beyond the electrochemical window of the electrolyte solvent.\cite{Janzakova2021b} In their corresponding Bode plots in Fig.~\ref{fig:fig4}g, we can observe that the same effect of fluctuating electrolyte resistance 1/\textit{G} exists at high frequencies, because also in this case, the number of branches is not easy to control with the growth. However, a discernible change in slope appears in the 1-10~kHz range, likely corresponding to the R|CPE$_3$ relaxation (pointing arrow in Fig.~\ref{fig:fig4}g). It is especially visible for duty-cycle values at 60\% and 40\%. While the high frequency plateau is usually flat and close to 0° in phase, in these specific extreme cases, the plateau appears to decrease in magnitude while the phase shows a slight bump above 0. The appearance of R|CPE$_3$ in the raw data of the EIS is attributed to the growth under an asymmetric duty-cycle. Using n~=~3 for Fig.~\ref{fig:fig1}e's model in these specific cases, similar values of $\alpha_1$ and $\alpha_2$ were obtained for all five samples in Fig.~\ref{fig:fig4}h. Only samples grown with duty-cycles of 40\%, 45\% and 60\% show convergence of the n=3 model for their impedance fitting, leading to a value for $\alpha_3$. 
The fitted time constants of the circuit appear to follow a symmetrical trend with the duty-cycle: the time constant $\tau_2$ gets longer as the system becomes more asymmetrical.
Considering $\alpha_3$ reaches a value below 0.5, it is unlikely that this R|CPE$_3$ relaxation should be modeled as a Voigt element, and a Randles element containing two constant phase elements could be physically more relevant here. In that case, the interpretation of the DRT could significantly change in Fig.~\ref{fig:fig4}i,j as the system would not be seen as a series of n Voigt elements anymore, making it not suitable if the diffusive term of the Randles element becomes dominant. We can however notice that if all relaxations follow Fig.~\ref{fig:fig1}e's model, we also observe a duty-cycle symmetry in Fig.~\ref{fig:fig4}j for $\tau_2$ and $\tau_3$ by the trajectory of their relaxation center (markers for 40\% and 60\% are closer the one another than to the other samples marker for these two relaxations).

\subsection{Discussion}
The structural evolution of conducting materials gathers natural peculiarities when compared to conventional electronics in which no mass transfer occurs. Various transports and charging modes govern CPD, from nucleation to interconnection\cite{Janzakova2021a,Kumar2022,Baron2024}. Their electrochemical relaxations become slower as they grow, while resistance takes over reactance in the impedance. Some relaxations are slow at the timescale of seconds: often seen as a drawback in conventional electronics where computing speed is key\cite{Seth2000,Markov2014}, such fading memory in the charged state can be relevant for unconventional computation where recurrence is an important ingredient of the analog information processing\cite{Vidal_Saez_2024,Kietzmann2019}. Multiple relaxations are simultaneously observed, but the most dominant is systematically the slowest one, characteristic of the dendrite structure, while the second one depends more on their environment. More relaxations are also evidenced, residual, but observable directly from the impedance raw data in some cases. The coexistence of multiple relaxations in a simple two electrodes system relies on the complexity of the interface dictating the dynamics. For instance, the polaron coupling between holes and anions intricates the slow dynamics of an electrolyte with the multiscale structuration of electrically conducting dendrites\cite{Paulsen2020,de_Levie_1963,Keddam_1984,Kaplan_1987,Jacquelin_1994}. This may explain the dispersive character of the measured impedances when dendrites do not connect. Despite new branches growing after the dendrites interconnected, their contribution in the system impedance is insignificant. Therefore, it appears necessary for a metamorphic electronics to operate in a regime where interconnectivity is only sparsely composed of completed dendrites to program the impedance between multiple nodes in a network. It is also noticed that the distance between the dendrites does not affect significantly the impedance of a junction, but a slower dynamics for the second relaxation is observed. This demonstrates the possibility to co-integrate multiple dendrites in a medium and to program the interconnectivity between them through their impedance, without their vicinity affecting the weight between specific interconnections. 
The relaxations' nature remains unclear: Nyquist plots do not feature Warburg elements with -45° straight lines, and the systematic observation of constant phase elements with $\alpha$ > 0.75 suggests the absence of diffusion limitation in the charging/discharging process. In contrast to multiple studies reporting diffusion limitation with PEDOT:PSS thin-films\cite{Wang_2023,Fabiano_2023, HernandezLabrado2011, Hern_ndez_Balaguera_2016, Nuramdhani2017}, the topological nature of such an electropolymerized amorphous material is likely to condition the impedance of the electrochemical system, rather than the crystallinity of a colloidal phase of conducting domains percolating in an ion-conducting ionomer\cite{Fabiano_2023, Paulsen2020}. Values of the attenuation coefficient $\alpha$ below 0.5 have also been observed: although they are not often seen experimentally with the exception of subdiffusion cases\cite{Metzler_2000}, they evidence important concerns on the nature of the charge transport through topological amorphous conductors. Various hypothesis can be proposed, such as the remain of polyelectrolyte excesses, trapped in the structure during the growth\cite{Ghazal2023}, or the difference in mobility of cations and polyanions in water’s bulk\cite{Koneshan_1998}, through water channels inside the dendrite\cite{Li2021b}, or through the dendrite material itself as a compound with the PSS cation conductor\cite{Reddy1971}. The symmetric DRT peaks suggest that the relaxations do not obey Gerischers’ impedance  underlying the Havriliak-Negami dispersion\cite{Boukamp_2020, Boukamp_2003}, but the possibility for multiple processes involving CPE and occurring in parallel should be considered. This can be explained by the existence of different ions or charged particles of various lengths having the same polarity but not the same permeation through a single dendrite morphology, and all occurring at a given time on the same dendritic electrode. As a possible way to verify these assumptions, a future characterization of these systems in the time domain shall be considered with different asymmetrical voltages to promote ion electrostatic attraction independently from diffusion with different morphologies of dendrites. The variation of their electrolyte environments under such stress shall also help decoupling the contributions of cations and anions, in addition to solvent properties which may play a key role in the signal transport, such as polarity, proton lability and viscosity.

\section{Conclusions}
In this study, we characterized electrogenerated PEDOT:PSS dendrites connecting two gold electrodes using EIS. An equivalent circuit model has been proposed based on the number of relaxations observed in the corresponding DRT. It has been shown that each observed relaxation has its own dependency to parameters such as the electrolyte composition and dendrites morphology. During their evolution, CPD go through three phases: the nucleation phase has a significant impact when the first monomers are electropolymerized, as a substantial drop in impedance magnitude occurs. The growth phase has an effect on the time constants of the primary relaxations which increase as more material is deposited and is branching out. Finally, at the connection phase, the conduction mode switches from mainly ionic to electronic, where most of the spectral features associated to the dendrite morphologies are lost in the impedance. A correlation has been identified between the dendrites' morphology and their corresponding impedance: the time constants of this system can be changed by tuning the thickness of the dendrite’s branches, but no specific link between individual parameters of each dendrite's constant phase elements and their morphology was established. By engraving the past-voltage history in their morphology, CPD interconnections are promising for future integration in unconventional computing devices, by storing electrically-accessible information with a low enthalpy process using carbon-based resources and additive manufacture. The perspective to implement it at the system level on a hardware classifier requires to verify the correspondence of the observed spectral features in the time domain. Future studies shall focus on programming such interconnects using spike voltage sequences in the time domain so as to observe the rich dynamics of these systems and the non-volatility of impedance modification through electropolymerization. This study paves the way for the implementation of recognition/classification tasks in future-emerging information-generation nodes embedding \textit{in materio} computing resources, and is also a step toward evolvable electronics shaping its interconnectivity in operando using both its environment and mass transfer as a way to self-manufacture and process electrical information in a circuit.

\newpage 

\section*{Acknowledgments}
The authors thank the French National Nanofabrication Network RENATECH for financial support of the IEMN cleanroom. We thank also the IEMN cleanroom staff for their advice and support. This work is funded by ANR-JCJC "Sensation" project (grant number: ANR-22-CE24-0001).

We would also like to thank Prof. Francesco Ciucci and Mr. Adeleke Maradesa for their guidance in selecting the parameters for the regularization involved in the DRT analysis.

\section*{Competing Interests}
The authors declare no competing interests.

\bibliographystyle{natsty-doilk-on-jour}  
\small
\bibliography{main}

\end{document}